# Proposal and Implementation of a novel perturb and observe algorithm using embedded software

Saad Motahhir, Abdelaziz El Ghzizal, Souad Sebti, Aziz Derouich

Laboratory of Production engineering, Energy and Sustainable Development (L.P.E.S.D.), higher school of technology, SMBA University Fez, Morocco

saad.motahhir@usmba.ac.ma, abdelaziz.elghzizal@usmba.ac.ma, souad_sebti@hotmail.com, aziz.Derouich@usmba.ac.ma

*Abstract*—The aim of this paper is to implement a modified Perturb and Observe algorithm (P&O), in order to solve the oscillation problem of photovoltaic (PV) output power generated by the conventional P&O algorithm. A comparison between the novel and the basic P&O algorithms is made. The first is implemented using embedded C language; the second is implemented using analog blocks. Next, the simulation study is made to present the response of the modified method to rapid temperature, solar irradiance, and load change.

## I. INTRODUCTION

Tracking the maximum power point (MPP) of a PV panel is an essential point of a PV system. It is decisive to operate the PV systems close to the MPP to improve its efficiency. Therefore, to optimize the energy withdrawn from the PV panel, we plan to insert a DC/DC converter controlled by an embedded system. This converter is implemented between the PV panel and the load. The embedded system is composed of two parts, the software, and the hardware. In this paper, we focus on the embedded software in which we implement a Maximum power point tracking (MPPT) algorithm.

For several years, research has been focusing on various MPPT control algorithms to draw the maximum power from the photovoltaic panel. Among these algorithms, there is perturb and observe (P&O) method with fixed and variable step size. This method, also known as the "hill climbing method", it presents oscillation around the MPP [10, 11]. Therefore a modified P&O algorithm is designed to minimize the oscillation in the output power.

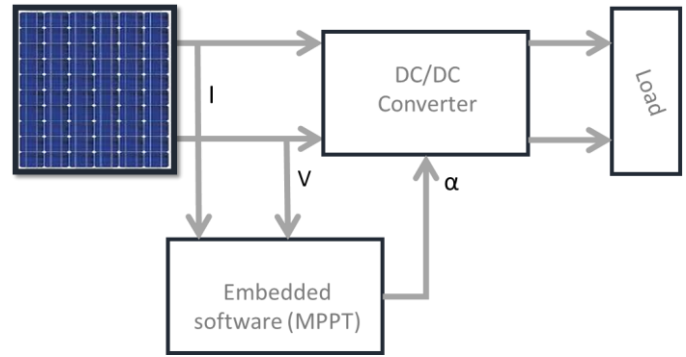

Figure 1. PV panel controlled by embedded software (MPPT)

The panel model is presented in Section I. Section II focuses on the implementation of the modified P&O algorithm using embedded C language, in order to compare its behavior to the basic P&O algorithm implemented using analog blocks. Moreover, the response of the modified method to rapid solar irradiance, temperature and load change is presented in this section.

## II. PANEL MODEL

A photovoltaic cell is a fundamental component in a photovoltaic panel. Since the net output voltage of a cell is very low, they are connected in parallel or in series or both ways, to meet practical demands. In order to mathematically model the PV cell, we derive the fundamental equation from the equivalent circuit of the solar cell shown in Fig. 2[1, 2, 3].

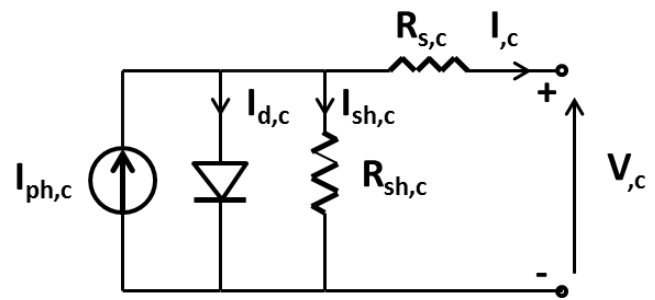

Figure 2. PV cell equivalent circuit

The current generated by the cell can be given as [7, 8, 9]:

$$I_{,c} = I_{ph,c} - I_{0,c}\left(\exp\frac{q(V_{,c} + R_{s,c} I_{,c})}{aKT} - 1\right) - \frac{(V_{,c} + R_{s,c} I_{,c})}{R_{sh,c}} \quad (1)$$

In this equation, $I_{ph,c}$ is the photocurrent, $I_{0,c}$ is the reverse saturation current of the diode, $V_{,c}$ is the voltage across the cell, q is the electron charge, **a** is the ideality factor of the diode, K is the Boltzmann's constant, T is the junction temperature, and $R_{s,c}$ and $R_{sh,c}$ are the series and shunt resistors of the cell, respectively.

Equation (1) of the elementary photovoltaic cell does not represent the I-V characteristic of a practical photovoltaic panel. Practical panels are composed of several connected photovoltaic cells. Cells connected in parallel increase the current and cells connected in series increase the output voltage.

In this study, we use MSX-60 panel. It is noted that the MSX-60, as several commercial PV panels, is constituted by two PV strings in series where each string is composed of 18 cells in series. The MSX-60 is modeled using PSIM software, and the simulation of the PV panel is presented in Fig.3.

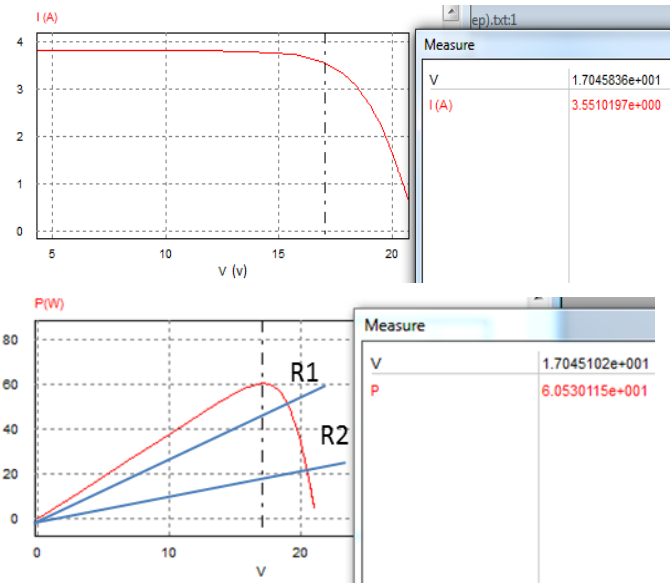

Figure 3. I-V and P-V curves for PV panel

As shown in P-V characteristic (Fig.3), the point where the power drawn by the load is the highest is called maximum power point (MPP), this power is depending on the temperature and irradiation, also the load imposes its own characteristic which is generally different from the MPP. In order to resolve this problem the MPPT command is used.

## III. MPPT IMPLEMENTATION USING EMBEDDED SOFTWARE

To operate the PV panel at the MPP, the DC/DC converter controlled by the MPPT is inserted between the PV panel and the load. There are several MPPT algorithms to locate the MPP. The most used algorithm is the Perturb and Observe (P&O) technique [4, 5, 6]. The reason lies in the fact that the P&O algorithm can be implemented in cheap digital devices ensuring high robustness and a good efficiency.

### A. DC/DC Converter

The power electronic converter used is a Boost converter inserted between the PV generator and the load [4]. It is characterized by its duty cycle α ($0 \leq \alpha \leq 1$) that gives the ratio between the input and the output voltage when the conduction is continuous (Fig. 4).

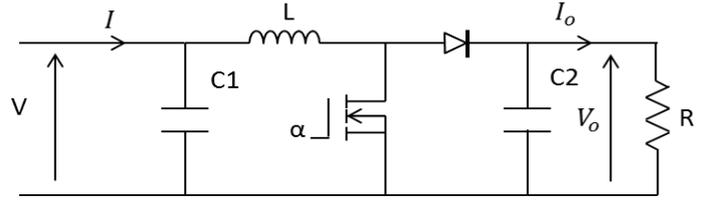

Figure 4. Boost converter

This converter operates by the following equations:

$$V_o = \frac{V}{(1-\alpha)} \quad (2)$$

$$I_o = I(1-\alpha) \quad (3)$$

Where, α, V and I, $V_o$ and $I_o$ are respectively the duty cycle, PV input voltage and current, the output voltage and current of the Boost converter.

The inductor value calculated to ensure the Boost converter operating in the continuous conduction mode is:

$$L \geq \frac{\alpha(1-\alpha)^2 R}{2F} \quad (4)$$

Where F is switching frequency and $\Delta I_L$ is peak-to-peak ripple of the inductor current.

The output capacitor value calculated to give the desired peak-to-peak output voltage ripple is:

$$C_O \geq \frac{\alpha I_o}{\Delta V_O F} \quad (5)$$

Where $\Delta V_O$ the output voltage peak-to-peak ripple.

## B. MPPT command

As a first step, the basic P&O algorithm is implemented using analog blocks (Fig. 5) and it is simulated in PSIM software as shown in Fig. 6. This method (P&O) generates oscillation in the output power; therefore its efficiency is not at the requested level (96.6%).

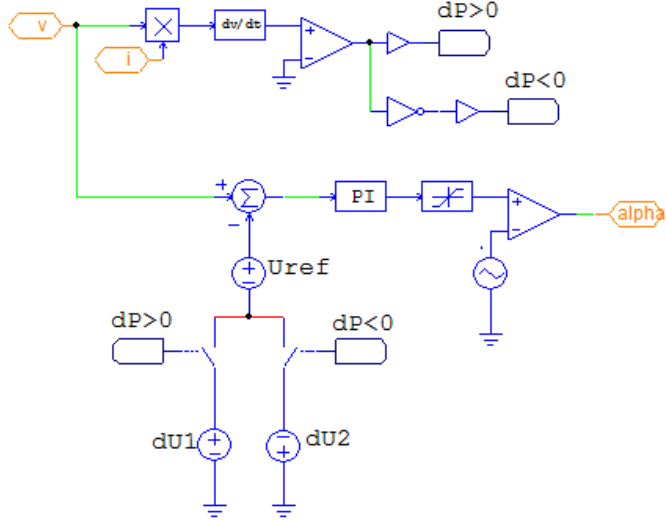

Figure 5. Basic P&O algorithm implemented by analog blocks

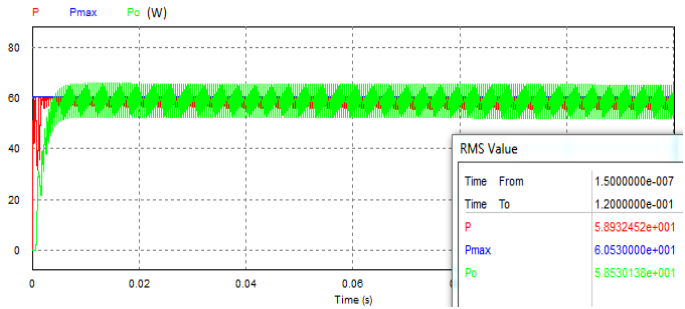

Figure 6. output Powers: P(Panel power), Pmax(Power max), Po(Load power)

So a modified P&O algorithm is designed in order to minimize the oscillation in the output power and it is implemented using C language (see Appendix). The reason lies in the fact that once the modified P&O algorithm is implemented using C language it can be implemented in a variety of cheap digital devices (microcontroller, DSP…) because C language is portable and machine-independent.

As shown in Fig. 7, the proposed MPPT is based on the conventional Perturb & Observe algorithm with two proposed additions: Firstly, the command of Boost converter alpha (α) will be modified only after every 5 periods. Secondly, when the power increases or decreases by a value lower than the epsilon threshold ε (where ε is considered as a small positive real number), α keeps its value, these additions can minimize oscillation in the output power, as well as the efficiency, will be improved.

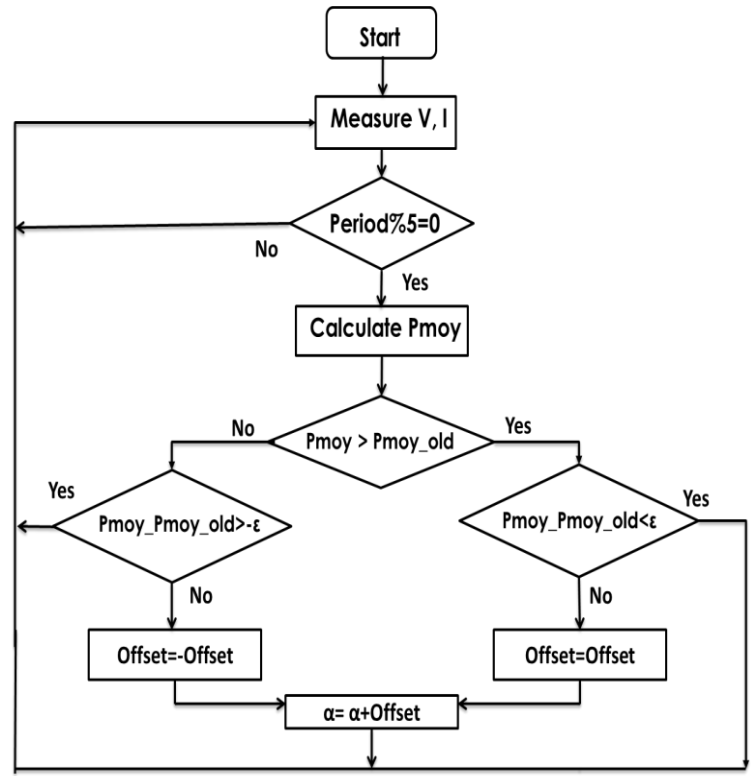

Figure 7. Flow process diagram of P&O control method for Boost

Fig. 8 describes the PV system implementing the proposed P&O algorithm using embedded C language in order to control the Boost converter, also the simulation result is depicted in fig. 9:

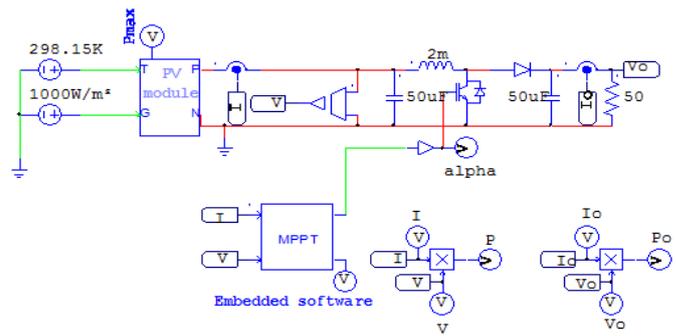

Figure 8. The PV generation system

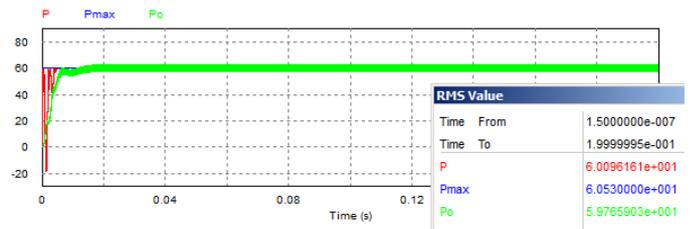

Figure 9. Output Powers: P(Panel power), Pmax(Power max), Po(Load power)

As it is shown in Fig. 9, using the modified algorithm, the oscillation in the output power are decreased. Also, our system takes just 0.005s to stabilize around the maximum power with an efficiency of ((Po/Pmax)*100=98.7376%).

The simulation study was made to illustrate the response of the proposed method to rapid temperature, solar irradiance, and load change. For this purpose, firstly the irradiation which is initially 200 W/m², is switched, at 0.09 s and 0.17 s, to 1000 W/m² and 800W/m² respectively, next, a brutal change from 800 to 500W/m² at 0.25s. Therefore, as shown in Fig. 10 the system is stable, even at brutal change, the system presents the oscillations which last less than 0.005s.

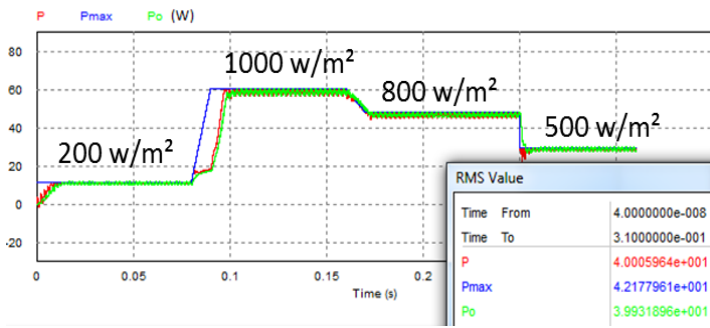

Figure 10. Power outputs: P(Panel power), Pmax(Power max), Po(Load power) for diffirents values of irradiation

Next, the temperature which is initially 25°C is switched at 0.06 s to 55°C. Therefore, as shown in Fig. 11 the increase in temperature causes a decrease in power; however, this decline lasts only 0.01s. After that, the embedded software backs the operating point and stabilizes it around the MPP.

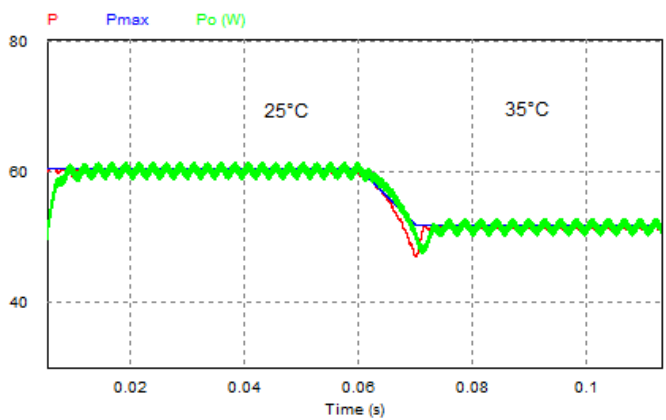

Figure 11. Power outputs: P(Panel power), Pmax(Power max), Po(Load power) for two values of temperature

Finally, in order to illustrate the response of the proposed method to the load change, we decrease the value of load from 50 Ω to 20 Ω, so as shown in Fig. 12, the decrease of load generates oscillation in the output power, however, and after 0.035 s the system converges to MPP.

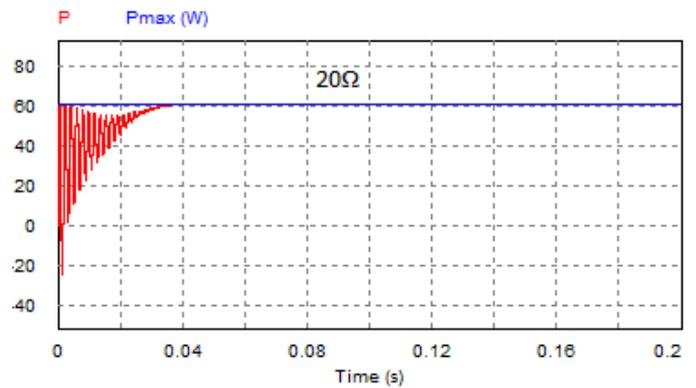

Figure 12. P(Panel power), Pmax(Power max) for a load of 20Ω

IV. CONCLUSION

A PSIM model simulation representing a PV panel is constructed. The panel is connected to the load through the Boost converter which is controlled by an embedded software. This embedded software presents a modified P&O algorithm developed by C language. As a result, a good efficiency is obtained compared to the basic P&O algorithm implemented by analog blocks. In addition, the modified method presents a good response to temperature, solar irradiance, and load change.

REFERENCES

[1] Tarak Salmi, Mounir Bouzguenda, Adel Gastli, Ahmed Masmoudi, " MATLAB/Simulink Based Modelling of Solar Photovoltaic Cell" ,INTERNATIONAL JOURNAL of RENEWABLE ENERGY RESEARCH, Vol.2, No.2, pp. 213-218, 2012.

[2] M. G. Villalva, J. R. Gazoli, E. Ruppert F., "MODELING AND CIRCUIT-BASED SIMULATION OF PHOTOVOLTAIC ARRAYS", Brazilian Journal of Power Electronics, Vol. 14, No. 1, pp. 35-45, 2009.

[3] M. G. Villalva, J. R. Gazoli, and E. R. Filho, "Comprehensive approach to modelling and simulation of photovoltaic arrays", IEEE Transaction on Power Electronics, Vol. 24, No. 5, pp. 1198-1208, 2009.

[4] Shridhar Sholapur, K. R. Mohan,T. R. Narsimhegowda, "Boost Converter Topology for PV System with Perturb And Observe MPPT Algorithm", IOSR Journal of Electrical and Electronics Engineering, Vol. 9, No. 4, pp. 50-56, 2014.

[5] Jaya Shukla , Jyoti Shrivastava, "Analysis of PV Array System with Buck-Boost Converter Using Perturb & Observe Method", INTERNATIONAL JOURNAL OF INNOVATIVE RESEARCH IN ELECTRICAL, ELECTRONICS, INSTRUMENTATION AND CONTROL ENGINEERING, Vol. 3, No. 3, pp. 51-57, 2015.

[6] Jubaer Ahmed, Zainal Salam, "An improved perturb and observe (P&O) maximum power point tracking (MPPT) algorithm for higher efficiency", Applied Energy, Vol. 150, pp. 97-108, 2015.

[7] Kamal Keshavani, Jigar Joshi, Vishrut Trivedi, Mitesh Bhavsar, "Modelling and Simulation of Photovoltaic Array Using Matlab/Simulink", INTERNATIONAL JOURNAL OF ENGINEERING DEVELOPMENT AND RESEARCH, Vol. 2, No. 4, pp.3742-3751, 2014.

[8] S. Rustemli, F. Dincer, "Modeling of Photovoltaic Panel and Examining Effects of Temperature in Matlab/Simulink", Electronics and Electrical Engineering, Vol. 109, No. 3, pp. 35-40, 2011.

[9] Savita Nema, R.K. Nema, Gayatri Agnihotri, "MATLAB/Simulink based study of photovoltaic cells/modules/array and their experimental verification", International journal of Energy and Environment, Vol. 1, No. 3, pp. 487-500, 2010.

[10] Ahmed M. Atallah, Almoataz Y. Abdelaziz, and Raihan S. Jumaahi, "Implementation of perturb and observe MPPT of PV system with direct control method using buck and buck-boost conveters", Emerging Trends in Electrical, Electronics & Instrumentation Engineering: An international Journal, Vol. 1, No. 1, pp. 31-44, 2014.

[11] C. Zhang, D. Zhao, J. Wang, G. Chen, "A modified MPPT method with variable perturbation step for photovoltaic system", Power Electronics and Motion Control Conference, pp. 2096-2099, 2009.

Please refer to the link below to download the data of this paper:

https://data.mendeley.com/datasets/m5yv5r2d6k/3

## APPENDIX

C block code of the modified P&O algorithm:

```c
#define NB_POINTS 1000
static int cpt =0,compt=0, cpt_periode = 0;
static int alpha =656, offset =3;
static float Panc =0, P =0, pmoyenne =0;
static double pacc =0;
out[1]=alpha;
// Generate PWM signal based on duty cycle 'alpha'
if(cpt<=alpha)
out[0] =1;
else
out[0] =0;
//Compute average value of PV power
if(cpt == NB_POINTS)
{
cpt =0;
compt ++;
pmoyenne = pacc/NB_POINTS ;
pacc =0;
//Duty cycle 'alpha' after each 5 periods by M-P&O algorithm
if(cpt_periode%5==0)
{
if(pmoyenne<Panc)
{
if(pmoyenne-Panc> -0.01)

else
{
offset=-offset;
alpha=alpha+offset;
}
}
else
{
if(pmoyenne-Panc < 0.01)
else
{offset=offset;
alpha=alpha+offset;}
}
if(alpha>=1000)
alpha=1000;
}
Panc = pmoyenne;
}
cpt++;
P=in[0]*in[1];
pacc=pacc + P ;
```